\documentclass[showpacs,pra]{revtex4}
\usepackage{amsfonts}
\usepackage{amsmath}
\usepackage{amssymb}
\usepackage{graphicx}

\begin{document}

\title{Dipolar spinor Bose-Einstein condensates}
\author{Su Yi$^1$ and Han Pu$^2$}
\affiliation{$^1$Institute of Theoretical Physics, Chinese Academy
of Sciences, Beijing, 100080, P.R. China}
\affiliation{$^2$Department of Physics and Astronomy, and Rice
Quantum Institute, Rice University, Houston, TX 77251, USA}

\begin{abstract}
Under many circumstances, the only important two-body interaction
between atoms in ultracold dilute atomic vapors is the
short-ranged isotropic $s$-wave collision. Recent studies have
shown, however, that situations may arise where the dipolar
interaction between atomic magnetic or electric dipole moments can
play a significant role. The long-range anisotropic nature of the
dipolar interaction greatly enriches the static and dynamic
properties of ultracold atoms. In the case of dipolar spinor
condensates, the interplay between the dipolar interaction and the
spin exchange interaction may lead to nontrivial spin textures.
Here we pay particular attention to the spin vortex state that is
analogous to the magnetic vortex found in thin magnetic films.
\end{abstract}

 \maketitle

\section{Introduction}

Experimental realization of dilute atomic Bose-Einstein
condensates (BECs) has revolutionalized the field of ultracold
atomic physics \cite{BEC1,BEC2,BEC3}. For the first time, we have
a macroscopic quantum object that is amenable not only to
exquisite experimental control, but also to detailed microscopic
theoretical description. Early experiments on atomic BECs were all
carried out in magnetic traps, where the atomic spin is polarized
by external magnetic fields, and hence the atomic spin degrees of
freedom is frozen (see, however, Refs.~\cite{jap,you}). The atoms,
however, get their spin degrees of freedom back when they are
trapped in off-resonant optical dipole traps \cite{spinor}, in
which case all magnetic Zeeman sublevels of the ground state atom
can be trapped. Such condensates are called spinor condensates.
Collisions between atoms give rise to an effective spin exchange
interaction \cite{ho2, ohmi}, which is analogous to the exchange
term in the theory of magnetism. The spin exchange interaction
leads to interesting coherent spin-mixing dynamics in spinor
condensates, a phenomenon that has been both theoretical studied
\cite{law,pu1,pu2, you1} and experimentally observed
\cite{stenger, jap1, seng, chap}.

Experimentally, spinor BECs have been realized in $^{23}$Na and
$^{87}$Rb. The ground state of both these atoms possesses a small
magnetic dipole moment of $\mu_B/2$, where $\mu_B$ is the Bohr
magneton. For typical condensate density ($\sim 10^{14}$
cm$^{-3}$), this would yield a tiny magnetic dipolar interaction
energy on the order of 0.1 nK per atom, which is a few orders of
magnitude smaller than the total collisional interaction energy.
Furthermore, it appears that the small dipolar strength would be
completely overwhelmed by any finite temperature effect (typical
temperatures in BEC experiments are $\sim 10$ nK). Consequently,
it has long been thought that dipolar interactions in such systems
can be safely ignored. Hence, experimental effort to achieve
dipolar atomic condensate has been focused on other types of
atoms, notably $^{52}$Cr \cite{pfau,pfau1} which has a magnetic
dipole moment of $6\mu_B$ in its ground state.

Under a more careful inspection, however, the conclusion that
dipolar interaction plays negligible role in alkali atoms becomes
questionable. We realized a few years ago that dipolar interaction
in spinor alkali condensates can play a more prominent role for the
following reasons:
\begin{enumerate}
\item Although the {\em total} collisional interaction strength is
much larger than the dipolar interaction strength, the spin
exchange interaction is not necessarily much stronger.
Particularly, for $f=1$ hyperfine manifold of $^{87}$Rb, the
dipolar energy can be as large as 10\% of spin exchange energy,
thus making a nontrivial contribution to the total {\em
spin-dependent} energy. \item The spin-dependent interaction,
although much weaker in magnitude than the spin-independent
interaction, is a critical determinant of the magnetic properties
of spinor condensates. \item The long-range and anisotropic nature
of the dipolar interaction may further enhance its effects. \item
A small finite temperature will not overwhelm the dipolar effects
due to the fact that, in a condensate, the interaction effects
enjoy a Bose-stimulation factor --- the total number of condensed
atoms. This same reasoning also explains the importance of the
weak nuclear dipolar interaction in superfluid state of $^3$He
\cite{he3}.
\end{enumerate}
This motivated us to carry out a detailed investigation of the
dipolar effects in spinor condensates. Our studies have confirmed
that rich physical phenomena can indeed be induced by the dipolar
interaction.

In the following, we will first present the Hamiltonian that
describes the system. We will then study the ground state
properties. To this end, two approaches will be used. The first is
the so-called single mode approximation --- all spin components of
the condensate are assumed to possess the same spatial wave
function. Under this approximation, the Hamiltonian can be greatly
simplified, which makes a full quantum mechanical study possible.
The single mode approximation, however, assumes that the atomic
spins are uniformly oriented in space. Its validity depends on the
dipolar interaction strength, as well as other parameters such as
the geometry of the trap potential. To go beyond the single mode
approximation, we adopt a second approach
--- the mean-field calculation without any {\em a priori}
assumption on the spatial wave functions. From this study we see
that sufficiently large dipolar strength induces non-trivial spin
textures in the ground state. The specific pattern of the spin
texture is sensitive to the trap geometry. In particular, a
pancaked-shaped trap favors the spin vortex state, analogous to
the magnetic vortices found in magnetic thin films or disks.

\section{Hamiltonian of a dipolar spinor condensate}
We consider $N$ condensed spin $f=1$ atoms trapped in an axially
symmetric harmonic potential,
\begin{equation}
V_{\rm ext}({\mathbf
r})=\frac{1}{2}M\omega^2(x^2+y^2+\lambda^2z^2)\,,\label{vext}
\end{equation}
with $\lambda$ being the trap aspect ratio, and $M$ the
atomic mass. We have chosen the symmetry axis to be the
quantization axis, $\hat{z}$. The atoms interact with each other
via both short-range collisions and long-range magnetic dipolar
interaction. Under a uniform magnetic field ${\mathbf B}$, the
second quantized Hamiltonian of the system reads
\[{\cal H} = {\cal H}_0 + {\cal H}_{\rm dd} \] where ${\cal H}_0$
and ${\cal H}_{\rm dd}$ represent the non-dipolar and dipolar part
of the Hamiltonian, respectively, and are given by
\begin{widetext}
\begin{eqnarray}
{\cal H}_0 &=&\int d{\mathbf r}\,\hat\psi_\alpha^\dag({\mathbf
r})\left[\left(-\frac{\hbar^2\nabla^2}{2M}+V_{\rm ext}({\mathbf
r})\right) \delta_{\alpha\beta}-g_F\mu_B{\mathbf B}\cdot{\mathbf
F}_{\alpha\beta}\right]\hat\psi_\beta({\mathbf
r})\nonumber\\
&&+\frac{c_0}{2}\int d{\mathbf r}\,\hat\psi_\alpha^\dag({\mathbf
r})\hat\psi_\beta^\dag({\mathbf r}) \hat\psi_\beta({\mathbf
r})\hat\psi_\alpha({\mathbf r})+\frac{c_2}{2}\int d{\mathbf
r}\,\hat\psi_\alpha^\dag({\mathbf r})
\hat\psi_{\alpha'}^\dag({\mathbf r}){\mathbf F}_{\alpha\beta}\cdot
{\mathbf F}_{\alpha'\beta'}\hat\psi_\beta({\mathbf
r})\hat\psi_{\beta'}({\mathbf r}),\label{hsp}\\
{\cal H}_{\rm dd} &=& \frac{c_d}{2}\int\int \frac{d{\mathbf
r}\,d{\mathbf r}'}{|{\mathbf r}-{\mathbf
r}'|^3}\left[\hat\psi_\alpha^\dag({\mathbf
r})\hat\psi_{\alpha'}^\dag({\mathbf r}'){\mathbf
F}_{\alpha\beta}\cdot{\mathbf
F}_{\alpha'\beta'}\hat\psi_\beta({\mathbf
r})\hat\psi_{\beta'}({\mathbf r}')
-3\hat\psi_\alpha^\dag({\mathbf
r})\hat\psi_{\alpha'}^\dag({\mathbf r'}) ({\mathbf
F}_{\alpha\beta}\cdot {\mathbf e})\,({\mathbf
F}_{\alpha'\beta'}\cdot{\mathbf e})\,\hat\psi_\beta({\mathbf
r})\hat\psi_{\beta'}({\mathbf r}')\right],\label{hspd}
\end{eqnarray}
\end{widetext}
where ${\mathbf F}$ is the spin angular momentum matrices,
${\mathbf e}=({\mathbf r}-{\mathbf r}')/|{\mathbf r}-{\mathbf
r}'|$ is a unit vector, and $\hat\psi_\alpha({\mathbf r})$ the
field operator for spin component (or Zeeman sublevel)
$\alpha=1,0,-1$. The collisional interaction parameters are
\cite{ho2,ohmi}
\[ c_0=\frac{4\pi\hbar^2(a_0+2a_2)}{3M}\,,\;\;\;
c_2=\frac{4\pi\hbar^2(a_2-a_0)}{3M}\,,\] where $a_F$ is the
scattering length for two $f=1$ atoms in the channel with total
spin angular momentum $F$. Symmetrization of many-body bosonic
wave function dictates that only the symmetric spin channels $F=0$
and 2 are involved. The dipolar interaction parameter is
\[c_d=\frac{\mu_0 g_F^2 \mu_B^2}{4\pi}\,,\] with $\mu_0$ being the
vacuum magnetic permeability, and $g_F$ the Land\'{e} g-factor.
Finally, in Eqs.~(\ref{hsp}) and (\ref{hspd}), and hereafter, it is
assumed that the repeated indices are summed over.

The first line of Eq.~(\ref{hsp}) represents the single-particle
part of the Hamiltonian, while the second line results from the
two-body contact interaction. The term proportional to $c_0$ is
symmetric in the spin indices and represents the spin-independent
contact interaction. The term proportional to $c_2$, on the other
hand, is spin-dependent and represents the short-range
spin-exchange interaction. The sign of $c_2$ determines the nature
of the spin-exchange coupling: negative $c_2$ represents
ferromagnetic coupling, while positive $c_2$ represents
antiferromagnetic coupling. The expression of ${\cal H}_{\rm dd}$
in Eq.~(\ref{hspd}) follows from the dipolar interaction potential
between two magnetic dipole moments ${\boldsymbol \mu}_i =g_F
\mu_B {\bf F}_i$ ($i=1,2$) located at spatial points ${\bf r}$ and
${\bf r}'$, respectively,
\begin{equation}
V_{\rm dd}({\bf r}, {\bf r}') =
\frac{\mu_0}{4\pi}\,\frac{{\boldsymbol \mu}_1 \cdot {\boldsymbol
\mu}_2 -3 ({\boldsymbol \mu}_1 \cdot {\mathbf e})\,({\boldsymbol
\mu}_2 \cdot {\mathbf e})}{|{\bf r}- {\bf r}'|^3} \,.\label{ddp}
\end{equation}

The spin-exchange term and the dipolar term describe two types of
spin-dependent interactions. It is the interplay and competition
between these two terms that give rise to a rich variety of spin
textures.

\section{Ground state under single mode approximation:
single-domain state}

\subsection{Hamiltonian under single mode
approximation}

The total Hamiltonian of the system as represented by
Eqs.~(\ref{hsp}) and (\ref{hspd}) is quite complicated. There
exists, however, a powerful method that can greatly simplify the
problem. This is the so-called single mode approximation (SMA). More
specifically, we assume that the field operators can be decomposed
as
\begin{equation}
\hat{\psi}_\alpha ({\bf r}) = \phi({\bf r})\, \hat{a}_\alpha
\,,\label{sma}
\end{equation}
where $\phi({\bf r})$ is a unit normalized {\em spin-independent}
spatial wave function. Here we shall not worry about the specific
expression of $\phi({\bf r})$, which should be properly chosen to
minimize the total energy.

Inserting Eq.~(\ref{sma}) into Eq.~(\ref{hsp}), we have
\begin{equation}
{\cal H}_0 =\int d{\mathbf r}\,\phi^*({\mathbf r})
\left(-\frac{\hbar^2\nabla^2}{2M}+V_{\rm ext}({\mathbf r})\right)
\phi({\bf r}) \, \hat{N} - g_F \mu_B {\bf B} \cdot \hat{\bf L} +
\frac{c_0}{2}\int d{\mathbf r}\,|\phi({\bf
r})|^4\,\hat{N}(\hat{N}-1) +\frac{c_2}{2}\int d{\mathbf
r}\,|\phi({\bf r})|^4\,(\hat{\bf L}^2-2\hat{N})\,,\label{h0sma}
\end{equation}
where $\hat{N} = \hat{a}_\alpha^\dag \hat{a}_\alpha$ is the total
particle number operator and $\hat{\bf L} = \hat{a}_\alpha^\dag
{\bf F}_{\alpha \beta} \hat{a}_\beta$ is the total spin angular
momentum operator. Following a similar procedure, we may obtain
${\cal H}_{\rm dd}$ under the SMA:
\begin{eqnarray}
{\cal H}_{\rm dd}&=&\frac{c_d}{2}\int d{\mathbf r}\int d{\mathbf
r}'\frac{|\phi({\mathbf r})|^2|\phi({\mathbf r}')|^2}{|{\mathbf
r}-{\mathbf r}'|^3}\left[\left(\hat{\bf L}^2-3(\hat{\mathbf
L}\cdot{\mathbf
e})^2\right)-\left(2\hat{N}-3\hat{a}_\alpha^\dag{\mathbf
F}_{\alpha\beta} \cdot{\mathbf e}\,{\mathbf
F}_{\beta\beta'}\cdot{\mathbf
e}\,\hat{a}_{\beta'}\right)\right]\nonumber\\
&=&\frac{c_d}{2}\int d{\mathbf r}\int d{\mathbf
r}'\frac{|\phi({\mathbf r})|^2|\phi({\mathbf r}')|^2}{|{\mathbf
r}-{\mathbf r}'|^3}\Big[\hat{L}_z^2(1-3\cos^2\theta_{\mathbf e})
-\frac{1}{4}(\hat{L}_+\hat{L}_-+\hat{L}_-\hat{L}_+)(1-3\cos^2\theta_{\mathbf e})\nonumber\\
&&\qquad-\frac{3}{2}(\hat{L}_+\hat{L}_z\cos\theta_{\mathbf
e}\sin\theta_{\mathbf e}e^{-i\varphi_{\mathbf e}}+h.c.)
-\frac{3}{2}(\hat{L}_-\hat{L}_z\cos\theta_{\mathbf
e}\sin\theta_{\mathbf e}e^{i\varphi_{\mathbf e}}+h.c.)
-\frac{3}{4}(\hat{L}_+^2\sin^2\theta_{\mathbf e}e^{-2i\varphi_{\mathbf e}}+h.c.)\nonumber\\
&&\qquad+\hat{a}_0^\dag \hat{a}_0(1-3\cos^2\theta_{\mathbf
e})-\frac{1}{2}(\hat{a}_1^\dag \hat{a}_1+\hat{a}_{-1}^\dag
\hat{a}_{-1})(1-3\cos^2\theta_{\mathbf e})
+\frac{3}{\sqrt{2}}\left(\cos\theta_{\mathbf e}\sin\theta_{\mathbf
e}e^{i\varphi_{\mathbf e}}\hat{a}_0^\dag
\hat{a}_1+h.c.\right)\nonumber\\
&&\qquad-\frac{3}{\sqrt{2}}\left(\cos\theta_{\mathbf
e}\sin\theta_{\mathbf e}e^{-i\varphi_{\mathbf e}}\hat{a}_0^\dag
\hat{a}_{-1}+h.c.\right)+\frac{3}{2}\left(\sin^2\theta_{\mathbf
e}e^{2i\varphi_{\mathbf e}}\hat{a}_{-1}^\dag
\hat{a}_1+h.c.\right)\Big]\,,\label{hddo}
\end{eqnarray}
where $\hat{L}_\pm\equiv \hat{L}_x\pm i\hat{L}_y$,
$\theta_{\mathbf e}$ and $\varphi_{\mathbf e}$ are the polar and
azimuthal angles of $({\mathbf r}-{\mathbf r}')$, respectively.
Obviously, this form of ${\cal H}_{\rm dd}$ is still quite
complicated. However, here we can take advantage of the spatial
symmetry of the system to further simplify the dipolar part. As we
have adopted an axially symmetric trapping potential, which is
indeed the case in most experiments, it is natural to assume that
the spatial wave function $\phi({\bf r})$ possesses the same axial
symmetry. Under this condition, it is not difficult to see that if
we carry out the integral in polar coordinates, terms proportional
to $e^{\pm im\varphi_{\mathbf e}}$ in Eq.~(\ref{hddo}) will not
survive after integrating over the azimuthal angle
$\varphi_{\mathbf e}$. We therefore have
\begin{equation}
{\cal H}_{\rm dd}=\frac{c_d}{4}\,\int d{\mathbf r} \int d{\mathbf
r}'\, |\phi({\mathbf r})\phi({\mathbf
r}')|^2\,\frac{1-3\cos^2\theta_{\mathbf e}}{|{\mathbf r}-{\mathbf
r}'|^3} \,(-\hat{\bf L}^2+3\hat{L}_z^2+3 \hat{n}_0 -\hat{N})\,,
\label{nn}
\end{equation}
where $\hat{n}_0 = \hat{a}^\dag_0 \hat{a}_0$ is the number
operator for spin-0 component.

The total Hamiltonian under the SMA is obtained by combining
Eqs.~(\ref{h0sma}) and (\ref{nn}). Since we are dealing with an
isolated system, the total number of atoms is a constant.
Therefore, we may neglect terms that only dependent on $\hat{N}$.
Finally, we have \cite{su1,su2}
\begin{equation}
{\cal H}_{\rm SMA}=(c_2'-c_d') \hat {\bf L}^2 +3c_d' (\hat L_z^2 +
\hat n_0)-g_F\mu_B {\mathbf B}\cdot \hat{{\mathbf L}}\,,
\label{qmham}
\end{equation}
where the two coefficients are defined as
\begin{equation}
c_2' =\frac{c_2}{2}\int d{\mathbf r}\,|\phi({\bf r})|^4\,,\;\;\;\;
c_d'=\frac{c_d}{4}\,\int d{\mathbf r} \int d{\mathbf r}'\,
|\phi({\mathbf r})\phi({\mathbf
r}')|^2\,\frac{1-3\cos^2\theta_{\mathbf e}}{|{\mathbf r}-{\mathbf
r}'|^3} \,. \label{coe}
\end{equation}

\begin{figure}[ptb]
\begin{center}
\includegraphics[
width=2.5in ]{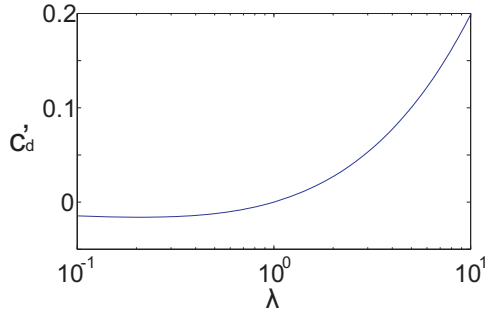}
\end{center}
\caption{The $\lambda$ dependence of $c_d'$ with $c_d=1$.}
\label{cdd}
\end{figure}

It is worth pointing out the trap geometry-dependence of the
effective dipolar coefficient $c_d'$. If we choose $\phi({\bf r})$
to be the single-particle ground state of the harmonic potential
$V_{\rm ext}$, i.e., \[ \phi({\bf r}) =
\lambda^{1/4}\pi^{-3/4}\,e^{-(x^2+y^2+\lambda z^2)/2}\,,\] where we
have used the harmonic oscillator length $\sqrt{\hbar/(M\omega)}$ to
be the units for length, the integral in $c_d'$ can be carried out
exactly:
\[ c_d' = \frac{c_d\sqrt{\lambda}}{6\sqrt{2\pi} (\lambda-1)}
\,\left(2\lambda +1 -3\lambda \frac{\tan^{-1}
\sqrt{\lambda-1}}{\sqrt{\lambda-1}} \right)\,.\] In Fig.~\ref{cdd},
we plot $c_d'$ as a function of trap aspect ratio $\lambda$. One can
see that the sign of $c_d'$ depends on the trap geometry:
\[ c_d' \left\{ \begin{array}{ll} < 0\,,& \;\;{\rm for \; a \;
prolate \; trap}\; (\lambda<1) \\ =0 \,, & \;\;{\rm for\; a\;
spherical\; trap}\; (\lambda=1) \\ >0\,, & \;\;{\rm for \; a \;
oblate \; trap}\; (\lambda>1) \end{array} \right. \] This provides
a convenient control knob one can use to change the properties of
or even induce phase transition in the system.

\subsection{Ground state structure under the single mode
approximation}
\begin{figure}[h]
\begin{center}
\includegraphics[
width=3.98in ]{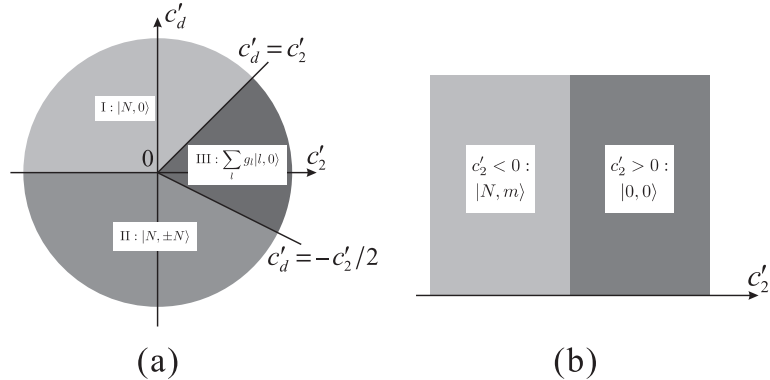}
\end{center}
\caption{Phase diagram of spinor condensate in the absence of
external magnetic fields. (a) Dipolar. (b) Non-dipolar.}
\label{domain}
\end{figure}

\begin{figure}[ptb]
\begin{center}
\includegraphics[width=3.in ]{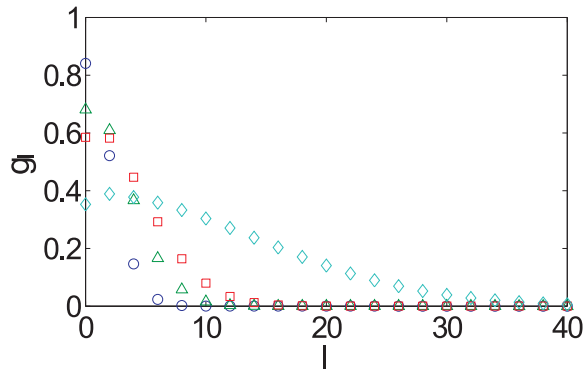}
\end{center}
\caption{The ground state in Region III can be written as
$\sum_l\,g_l|l,0\rangle$. Here we plot the coefficients $g_l$ for
$c_d'/c_2'=0.2$ (circle), $0.8$ (triangle), $1.4$ (square), and
$2.0$ (diamond). We have taken $N=100$.} \label{gl}
\end{figure}

Hamiltonian (\ref{qmham}) is reminiscent of the Hamiltonian that
describes a quantum magnet. The ground state can be obtained by
diagonalizing (\ref{qmham}). In the absence of the external
magnetic field, i.e., ${\bf B}=0$, the phase diagram can be
plotted in the $c_2'-c_d'$ parameter space as shown in
Fig.~\ref{domain}(a). According to the nature of the ground state,
we can divide the parameter space into three regions labelled as
I, II and III. Region I represents a ferromagnetic phase with
easy-plane anisotropy. The ground state wave function in this
region can be written as $|N,0\rangle$, where we have used the
standard angular momentum basis state $|l,m\rangle$ such that \[
\hat{\bf L}^2 |l,m\rangle = l(l+1)\hbar^2\,
|l,m\rangle\,,\;\;\;\hat{L}_z |l,m\rangle =
m\hbar\,|l,m\rangle\,.\]Region II represents a ferromagnetic phase
with easy-axis anisotropy. The ground state wave function in this
region can be written as $|N,\pm N\rangle$ with a two-fold
degeneracy. In both Regions I and II, the atomic spins are aligned
along the same direction, either in the transverse plane (Region
I) or along the $z$-axis (Region II). Region III, on the other
hand, represents roughly an anti-ferromagnetic phase where atomic
spins are entangled to form spin singlets. Here the ground state
wave function has a slightly more complicated form: $\sum_l g_l
|l,0\rangle$ where the coefficients $g_l$ has in general to be
calculated numerically. Several examples are given in
Fig.~\ref{gl}. The mixing of different $l$ states is due to the
$\hat{n}_0$ term in the Hamiltonian. In principle, the $\hat{n}_0$
term will have a similar effect in the other two regions I and II.
However, its effect there is negligible for large particle numbers
$N \gg 1$.

For comparison, we also present the corresponding phase diagram
for the non-dipolar case ($c_d'=0$) \cite{law} in
Fig.~\ref{domain}(b). Under this situation, the Hamiltonian is
simply ${\cal H} = c_2' \hat{\bf L}^2$ which possesses a full
rotational symmetry in spin space. The ground state is determined
by the sign of $c_2'$. It is an isotropic Heisenberg
(anti-)ferromagnet if $c_2'<0$ ($c_2'>0$). The effect of the
dipolar interaction is therefore quite transparent: It breaks the
rotation symmetry of the non-dipolar system and introduces
magnetic anisotropy.

Next we investigate the effect of a uniform external field. In
particular, we are interested in the critical field strength at
which the system is fully polarized by the external field.

{\em Longitudinal field} --- First consider a longitudinal field
along the $z$-axis. It is easy to see that under this condition,
$\hat{L}_z$ is still a constant of motion since it commutes with the
Hamiltonian (\ref{qmham}). Its effect in Region II is quite obvious:
Any longitudinal field will break the degeneracy of the ground state
in the absence of the external field and polarize the spins along
the field. Therefore the critical field strength here is
infinitesimally small. For Region I, the new ground state should
have the form $|N, m\rangle$ where the value of $m$ may be obtained
by minimizing the energy \[ E(m) = \langle N,m|{\cal H}_{\rm SMA}
|N,m \rangle \,,\] which yields \[ m = \left[ \frac{g_F \mu_B
B}{c_d'} + \frac{1}{2} \right]\,,\] where $[x]$ denotes the largest
integer no larger than $x$. Critical field strength is reached when
$m=N$ or
\[B_c = \frac{6c_d' (N-1/2)} {g_F \mu_B}\,.\] For Rb condensate,
this would correspond to a field strength on the order of 0.1 mG.
The critical field strength for Region III can be obtain in a
similar manner. Here the ground state has the form $\sum_{l\le m_0}
g_l |l,m_0\rangle$. The critical field is given by \[ B_c =
\frac{(4N-5)c_d' + (2N-1)c_2'}{g_F \mu_B} \,.\]

\begin{figure}[ptb]
\begin{center}
\includegraphics[
width=3.98in ]{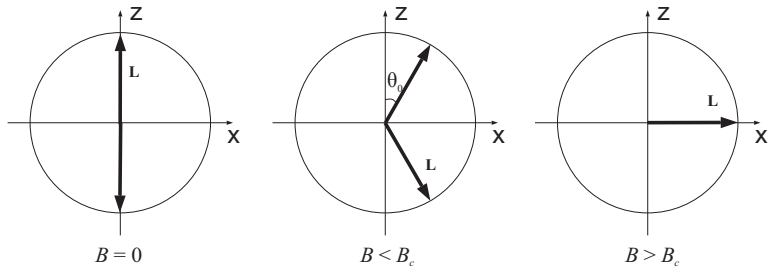}
\end{center}
\caption{The effect of transverse field in Region II. Classical
picture of the spin orientation. The ground state is doubly
degenerate when $B<B_c$ and fully polarized when $B \le B_c$.}
\label{spin}
\end{figure}
{\em Transverse field} --- Next we consider a transverse field
along, e.g., the $x$-axis. Its effect in Region I is similar to that
of a longitudinal field in Region II: any transverse field would
fully polarize the system in Region I. For Region II, we can take a
classical approach since here the total spin is macroscopically
large ($L \approx N \gg 1$). Treat the spin as a classical
magnetization vector with length $L=N$ and polar angle $\vartheta$,
we can write down the energy of the system as (neglecting the
unimportant $\hat{n}_0$-term)
\[ E(\vartheta) = (c_2'-c_d')N^2 + 3c_d' N^2 \cos^2 \vartheta -
g_F\mu_B B N \sin \vartheta \,.\] Minimizing the energy with
respect to $\vartheta$ yields the optimal value for $\vartheta$ as
\[ \vartheta_0 = \frac{\pi}{2} \pm \left( \frac{\pi}{2} +
\sin^{-1} \frac{g_F \mu_B B}{6c_d' N} \right) \,.\] Hence the
critical field strength is \[B_c = \frac{6c_d' N}{g_F \mu_B} \,.\]
For $B<B_c$, the ground state is doubly degenerate as
schematically shown in Fig.~\ref{spin}. When $B \ge B_c$, the two
degenerate states collapse into one and the system is fully
polarized by the transverse field. We want to point out that, in a
quantum mechanical treatment, the double degeneracy at $0<B<B_c$
will be lifted by quantum fluctuations, resulting in quantum spin
tunnelling \cite{tul}. Such quantum effects will be important when
$N$ is small.

Finally, for Region III, the ground state can again only be obtained
numerically as a superposition of different angular momentum states.
The critical field can be shown to be \[ B_c
=\frac{2(c_2'-c_d')N}{g_F \mu_B} \,.\]

\subsection{Validity of single mode approximation}
\label{SMA} Our discussion so far is constrained within the SMA. As
can be seen, the SMA is a powerful approximation from which rich
physics can be derived. Now we want to address the important
question of the validity of this approximation.

First we notice that the SMA implies that the spin orientation is
position independent. To show this, we define the normalized spin
vector as \[ {\bf s}({\bf r}) =  {\bf S}({\bf r}) /n({\bf r})\,,\]
where $n({\bf r})$ is the number density, and ${\bf S}({\bf r})
\equiv \langle \hat{\psi}_\alpha^\dag ({\bf r}) {\bf F}_{\alpha
\beta} \hat{\psi}_\beta ({\bf r})\rangle$ is the total spin
density. Under the SMA, we have $n({\bf r}) = N|\phi ({\bf
r})|^2$, and ${\bf S}({\bf r}) = n({\bf r})\,\langle
\hat{a}_\alpha^\dag {\bf F}_{\alpha \beta} \hat{a}_\beta \rangle =
n ({\bf r}) \langle \hat{\bf L} \rangle$. Therefore, we have ${\bf
s} =\langle \hat{\bf L} \rangle$ which is spatially invariant.
This situation would correspond to the so-called {\em
single-domain} state of a nanomagnetic material \cite{magbook}.
The spatially uniform spin orientation minimizes the short-range
spin-exchange interaction, but does not in general minimize the
dipolar energy which is most effectively reduced for closure
domain structures. Obviously the single-domain state is thus
stable when the exchange interaction dominates. When this is not
the case, the exchange interaction will be frustrated by the
dipolar interaction, rendering the single-domain state
intrinsically unstable to various other magnetically ordered
states. Armed with this insight, we conclude that the SMA is valid
only when the dipolar strength is sufficiently small and hence
cannot cover the whole spectrum of interesting quantum spin
phenomena.

Our study confirms this conclusion. For large dipolar strength,
the SMA is no longer valid and nontrivial spatial spin textures
develop in the system. We shall now turn to this situation that
goes beyond the SMA.

\section{Beyond Single Mode Approximation: Spin Textures}
\subsection{Mean-Field Equations}
To go beyond the SMA, we need to calculate the wave functions for
each spin component explicitly. To this end, we adopt the standard
mean-field approach. Specifically, the field operators
$\hat{\psi}_\alpha$ in (\ref{hsp}) and (\ref{hspd}) are replaced
by their corresponding expectation values $\psi_\alpha = \langle
\hat{\psi}_\alpha \rangle$. The Hamiltonian thus becomes the
energy functional:
\begin{eqnarray}
E[\psi_\alpha, \psi_\alpha^*] &=&\int d{\mathbf
r}\left[\frac{\hbar^2
(\nabla\psi^*_\alpha)(\nabla\psi_\alpha)}{2m} +V_{\rm ext}({\bf
r})n({\bf r}) - g_F\mu_F {\bf B}\cdot{\bf
S}({\bf r})+\frac{c_0n^2({\bf r})}{2}+\frac{c_2{\mathbf S}^2({\bf r})}{2}\right] \nonumber \\
&& +\frac{c_d}{2}\int \frac{d{\mathbf r}d{\mathbf r}'}{|{\bf
r}-{\bf r}'|^3}\left[{\mathbf S}({\mathbf r})\cdot{\mathbf
S}({\mathbf r}')-3\left({\mathbf S}({\mathbf r})\cdot{\mathbf
e}\right)\,\left({\mathbf S}({\mathbf r}')\cdot{\mathbf
e}\right)\right]\,, \label{energy}
\end{eqnarray}
where $n({\bf r})=\psi_\alpha^*({\bf r})\psi_\alpha({\bf r})$ and
${\mathbf S}({\bf r})=\psi^*_\alpha{\mathbf
F}_{\alpha\beta}\psi_\beta$ are the total number density and spin
density, respectively. Minimization of the energy functional
$E[\psi_\alpha,\psi_\alpha^*]$ requires $\delta
E[\psi_\alpha,\psi_\alpha^*]/\delta \psi^*_\alpha =0$, which
yields a set of nonlinear coupled partial differential equations
for the condensate wave function $\psi_\alpha({\bf r})$:
\begin{eqnarray}
i\hbar \frac{\partial \psi_1}{\partial t} &=& \left[
-\frac{\hbar^2 \nabla^2}{2M} +V_{\rm ext}- g_F\mu_F {\bf
B}\cdot{\bf F}_{1\beta} \psi_\beta + c_0n+ c_2 S_z + c_d {\cal
D}_z \right] \psi_1 + [c_2 S_-  +c_d {\cal
D}_-]\psi_0 \,, \\
i\hbar \frac{\partial \psi_0}{\partial t} &=& \left[
-\frac{\hbar^2 \nabla^2}{2M} +V_{\rm ext}- g_F\mu_F {\bf
B}\cdot{\bf F}_{0\beta} \psi_\beta +c_0 n \right]\psi_0 + [c_2 S_+
+ c_d {\cal D}_+ ] \psi_1 + [c_2 S_- + c_d
{\cal D}_- ] \psi_{-1}\,,  \\
i\hbar \frac{\partial \psi_{-1}}{\partial t} &=& \left[
-\frac{\hbar^2 \nabla^2}{2M} +V_{\rm ext}({\bf r})- g_F\mu_F {\bf
B}\cdot{\bf F}_{-1\beta} \psi_\beta + c_0n- c_2 S_z - c_d {\cal
D}_z \right] \psi_{-1} + [c_2 S_+  +c_d {\cal D}_+]\psi_0 \,,
\end{eqnarray}
where $S_{\pm} \equiv (S_x \pm i S_y)/\sqrt{2}$ and the integral
operator ${\cal D}_i$ is given by
\[{\cal D}_i ({\bf r}) = \int d{\bf r}'\, \frac{1}{|{\bf r}-{\bf
r}'|^3}\,\left[ S_i ({\bf r}') - 3e_i \,{\bf S}({\bf r}')\cdot
{\bf e} \right]\,. \] The ground wave functions are obtained by
solving these equations using the imaginary-time evolution method,
where the term involving the integral operator ${\cal D}_i$ are
dealt with using convolution theorem and fast Fourier transform.

\subsection{Phase Diagram}
\begin{figure}[h]
\begin{center}
\includegraphics[
width=3.98in ]{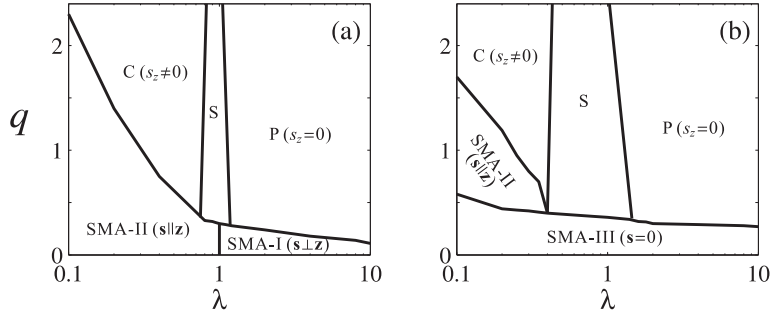}
\end{center}
\caption{Phase diagram of a dipolar spin-1 condensate with (a)
ferromagnetic ($c_2<0$) and (b) anti-ferromagnetic ($c_2>0$) spin
exchange coupling. Regions labelled by SMA-I, II, and III are
where the SMA is shown to be valid.} \label{diagram}
\end{figure}
From this calculation, the validity of the SMA can be directly
checked by comparing the wave functions $\psi_\alpha$. Phase
diagrams such as shown in Fig.~\ref{diagram} can be obtained
\cite{su3,ueda}. Fig.~\ref{diagram}(a) is the phase diagram for a
ferromagnetic exchange coupling with $c_2<0$ in the absence of
external magnetic field. Here we fix the ratio $c_2/c_0$ to be
$-0.01$ corresponding to the values of $^{87}$Rb. The phase
diagram is plotted in the parameter space spanned by the trap
aspect ratio $\lambda$ and the ratio between the dipolar and the
exchange interaction strength $q\equiv c_d/|c_2|$. A similar phase
diagram for an anti-ferromagnetic exchange coupling ($c_2>0$) is
plotted in Fig.~\ref{diagram}(b). Here we use $c_2/c_0=0.03$ as in
the case of $^{23}$Na.

From Fig.~\ref{diagram}, we can make the following observation:
\begin{enumerate}
\item For any given trap geometry, the SMA is only valid for
sufficiently small dipolar strength, in full agreement with the
qualitative argument we made in Sec.~\ref{SMA}. \item In the region
where the SMA is valid, according to the nature of the wave function
which determines the local spin vector ${\bf s}$, we can further
divide the region into three subregions labelled as SMA-I, II and
III. They correspond to Regions I, II and III shown in
Fig.~\ref{domain}. \item The critical value of $q$ where the SMA
becomes invalid is sensitive to the trap aspect ratio: it decreases
as $\lambda$ increases, i.e., when the trap becomes more and more
pancake-shaped. This shows that the dipolar interaction plays a more
prominent role in a pancake geometry, which is consistent with the
observation in the study of magnetic materials
--- The dipolar interaction, normally weak enough to be ignored in
bulk materials, plays an essential role in stabilizing long-range
magnetic order in two-dimensional systems such as magnetic thin
film \cite{thin}. \item In the region where the SMA is invalid, we
can also divide the region into three subregions labelled C, S and
P. We will discuss each of these regions in more detail below.
\end{enumerate}

\subsection{Ground State Beyond Single Mode Approximation}
\begin{figure}[h]
\begin{center}
\includegraphics[
width=5in ]{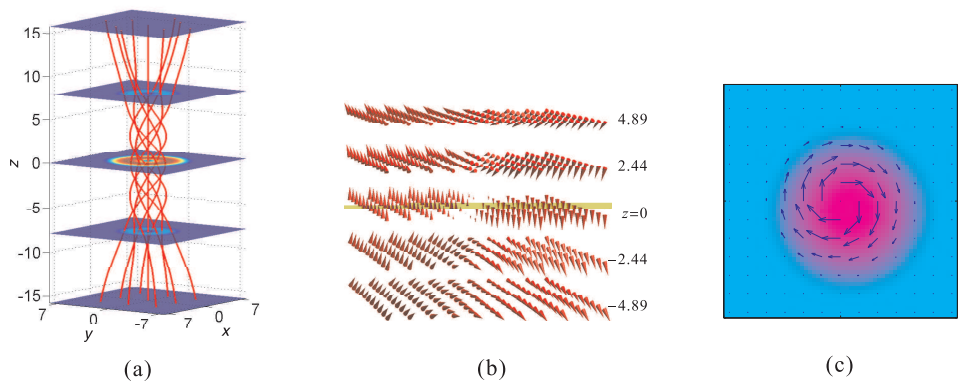}
\end{center}
\caption{Ground state spin structure in (from left to right) C, S
and P regions. (a) is a streamline plot of the local spins in a
cigar-shaped condensate. Here the spins have a dominant
$z$-component and form a helical pattern. In (b), the arrows
indicate the direction of the local spin vector. (c) is the spin
structure in the $z=0$ plane for a pancake-shaped condensate.
Here, the spins are planar and lie in the $x-y$ plane. Spins at
different layers of $z$ have the same vortex structures.}
\label{texture}
\end{figure}
Now let us discuss in more detail the ground state structure of
the spinor condensate in the region where the SMA becomes invalid.
The phase diagram plotted in Fig.~\ref{diagram} shows that there
are three regions labelled as C, S and P, which stands for cigar,
spherical and pancake, indicating the geometry of the trap in
which a particular type of spin texture is favored.

\begin{figure}[ptb]
\begin{center}
\includegraphics[
width=4.in ]{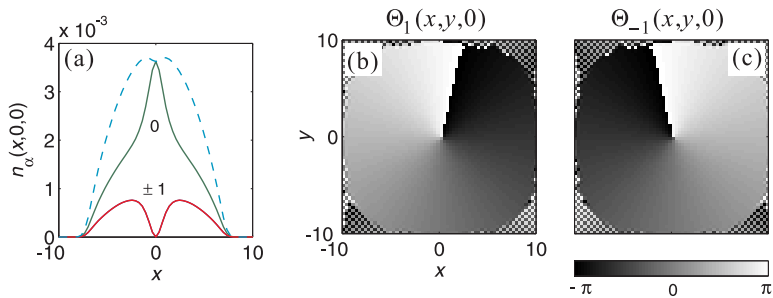}
\end{center}
\caption{(a) Density of each each spin component along the
$x$-axis (solid lines). The total density is represented by the
dashed line. (b) and (c) illustrate the phases $\Theta_1$ and
$\Theta_{-1}$ in the $z=0$ plane.} \label{vortex}
\end{figure}

Figure~\ref{texture} illustrates the ground state spin structure
in these three regions. In a pancake-shaped trap (Region P of
Fig.~\ref{diagram}), spins are planar, i.e., lying in the
transverse plane. Dipolar interaction forces the atomic spin to
form a vortex pattern. We can write the wave function as
$\psi_\alpha ({\bf r}) = \sqrt{n_\alpha({\bf r})}\,
e^{i\Theta_\alpha ({\bf r})}$. Our calculations show that in this
region, the wave function contains a nontrivial phase:
\begin{equation}
\Theta_\alpha ({\bf r}) = w_\alpha \varphi + \phi_\alpha \,,
\label{phase}
\end{equation}
where $\phi_\alpha$ is a constant phase shift whose absolute value
is not important but satisfies $\phi_1+ \phi_{-1} - 2\phi_0=0$,
$\varphi$ is the azimuthal angle, and $w_\alpha$ is then the
``winding number''
--- circling once around the $z$-axis, the phase of the wave
function $\psi_\alpha$ changes by $w_\alpha \times 2\pi$. The
specific values for the winding numbers are
\[\langle w_1,w_0,w_{-1}\rangle=\langle-1,0,1\rangle\,.\]
Furthermore, we also have $n_1({\bf r}) = n_{-1}({\bf r})$.
Therefore, we can see that the spin component 1 and $-1$ share the
same density, but opposite winding number as illustrated in
Fig.~\ref{vortex}. The spin vortex state is obviously analogous to
the magnetic vortex state found in magnetic thin films
\cite{magbook,thin}. However, there are important differences. In
the central core region of a magnetic vortex, due to local
magnetization conservation, the magnetic moment has to align
perpendicular to the plane of the film in order to decrease the
exchange energy at the center of the vortex. In a spinor
condensate, however, there is no such constraint on local spin
moments. Therefore local spin can simply vanish in the vortex
core. Indeed, the core region is occupied by the spin-0 component
while the other two spin components vanish. The spin vector can be
expressed as \[{\bf S}({\bf r})=\left(
\begin{array}{lll} S_x \\ S_y \\ S_z \end{array} \right) = 2\sqrt{2n_0 n_1}\,
\left(\begin{array}{ccc} \sin \varphi \\ -\cos \varphi\\ 0
\end{array} \right)\,.\] Such a spin texture is also referred to as a coreless
skyrmion. Further, the spinor condensate is really a novel
superfluid described by macroscopic wave functions which contain
both amplitude and phase. In quantum mechanics, the spatial
non-uniformity of the phase of a single-particle wave function is
directly related to the velocity of the particle as
\[{\bf v}_\alpha = \frac{\hbar}{M} \nabla \Theta_\alpha \,.\] We
therefore conclude that, in the spin vortex state, spin components
1 and $-1$ circle around the $z$-axis in opposite directions,
while the spin-0 component is stationary as $w_0=0$. This results
in a situation with a net spin current but without a mass current.

In a cigar-shaped trap (Region C of Fig.~\ref{diagram}), local
atomic spins are predominantly aligned along the $z$-axis. In that
way, most of the spins are aligned in a head-to-tail configuration
which minimizes the dipolar energy. Here the phase of the wave
functions can still be expressed as in Eq.~(\ref{phase}) but with
a different set of winding numbers \[\langle
w_1,w_0,w_{-1}\rangle=\langle 0,1,2 \rangle\,.\] Further the phase
shift $\phi_\alpha$ are no longer constants but functions of $z$.
The spin vector in the C phase takes the form
\begin{eqnarray}
{\mathbf S}({\bf r})=\left(\begin{array}{ccc} \Delta
\sin(\varphi+\delta) \\ -\Delta \cos(\varphi+\delta)\\
n_1-n_{-1} \end{array} \right)\,,\nonumber
\end{eqnarray}
where $\Delta \equiv \sqrt{2n_0}(\sqrt{n_1}+\sqrt{n_{-1}})$, and
$\delta(z)\equiv\varphi_0(z)-\varphi_1(z)$ is the spin twisting
angle which is a monotonically increasing function of $z$ with
$\delta(z=0)=0$.

In between the P and C phases, is the S phase which occurs when
the trapping potential is close to spherical ($\lambda \approx
1$). A distinct feature of the S phase is that $n_\alpha$ becomes
non-axisymmetric, signalling the breakdown of the cylindrical
symmetry of the spatial wave functions. A typical spin
configuration in S phase is displayed in the middle plot in
Fig.~\ref{texture}. One can clearly see that the spin structure
does not possess the cylindrical symmetry and there is a
180$^\circ$-domain wall separates two spatial regions of spin.

\subsection{Spin Vortex in External Magnetic Field}
We have seen now that the dipolar interaction plays a more
important role in pancake-shaped traps in which the ground state
takes the form of a spin vortex. In this section we will focus on
the spin vortex state and study how it responses to an external
magnetic field.

\begin{figure}[ptb]
\begin{center}
\includegraphics[
width=5.in ]{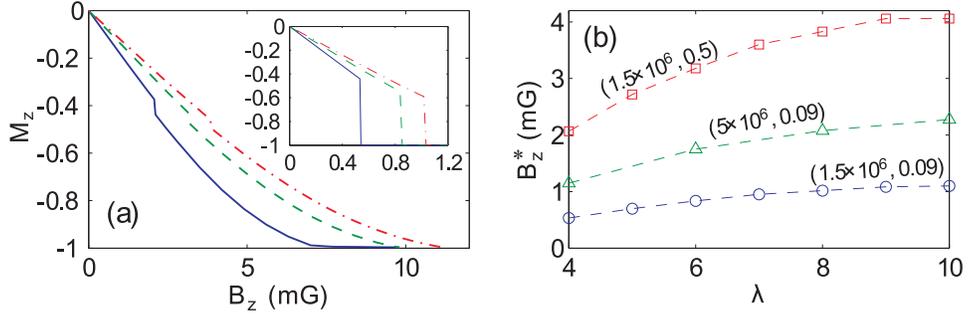}
\end{center}
\caption{(a) The B-field dependence of magnetization $M_z$ for
$q=0.5$, $N=1.5 \times 10^6$ and $\lambda=4$ (solid line), $6$
(dashed line), and $8$ (dash-dotted line). The inset shows the
same thing except for $q=0.09$ which corresponds to the real
parameters for $^{87}$Rb. (b) The $\lambda$ dependence of critical
magnetic field for various $(N,q)$ values. } \label{mvorbz}
\end{figure}

{\em Longitudinal field} --- First consider a uniform magnetic
field along the $z$-axis. Here we define the magnetization as \[
M_z = \frac{1}{N} \,\int d{\bf r}\,S_z({\bf r})=\frac{1}{N} \,\int
d{\bf r}\,[n_1({\bf r})-n_{-1}({\bf r})]\,.\]  We plot in
Fig.~\ref{mvorbz}(a) the magnetization as a function of the field
strength $B_z$. As expected, the magnitude of $M_z$ is a
monotonically increasing function of $B_z$ ($M_z$ is negative
here, because for the $f=1$ hyperfine level of alkali atoms,
$g_F=-1/2<0$). However, there exists a critical field strength
$B^*_z$ at which $M_z$ suddenly jumps, signalling a first-order
phase transition of the system. This critical field strength
depends on the trap aspect ratio, the total number of atoms and
the dipolar strength, as shown in Fig.~\ref{mvorbz}(b).

\begin{figure}[h]
\begin{center}
\includegraphics[
width=2.5in ]{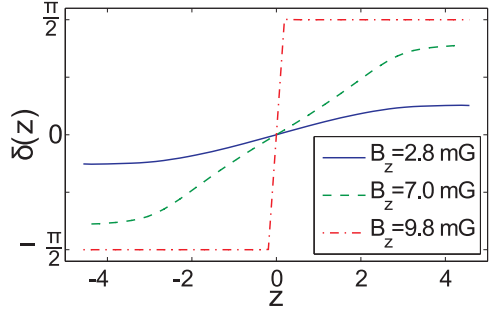}
\end{center}
\caption{ The $z$ (in units of $\sqrt{\hbar/M\omega}$) dependence
of spin twisting angle $\delta(z) = \phi_1(z)-\phi_0(z)-\pi/2$ for
$q=0.5$ and $N=1.5 \times 10^6$.} \label{delta}
\end{figure}
To provide more insights into this phase transition, we study the
wave functions $\psi_\alpha ({\bf r}) = \sqrt{n_\alpha({\bf r})}\,
e^{i\Theta_\alpha ({\bf r})}$. We find that the phases
$\Theta_\alpha$ can still be written as in Eq.~(\ref{phase}). The
phase shifts $\phi_\alpha$ always satisfy the condition
$\phi_1+\phi_{-1}-2\phi_0=0$. However, for $B_z<B^*_z$,
$\phi_\alpha$ are constants; while for $B_z>B^*_z$, $\phi_\alpha$
become functions of $z$ as shown in Fig.~\ref{delta}. Furthermore,
the winding numbers also change across the critical field and are
given by
\[\langle w_1,w_0,w_{-1}\rangle= \left\{ \begin{array}{ll}
\langle-1,0,1\rangle \,,& B<B^* \\ \langle-2,-1,0\rangle \,,&
B>B^*
\end{array} \right.\] The local spin vector takes the form
\[{\bf S}({\bf r}) = \left( \begin{array}{ccc}
\sqrt{2n_0}(\sqrt{n_1}+\sqrt{n_{-1}})\,\sin(\varphi  -\delta) \\
-\sqrt{2n_0}(\sqrt{n_1}+\sqrt{n_{-1}})\,\cos(\varphi  -\delta) \\
n_1-n_{-1} \end{array} \right) \] with $\delta =
\phi_1-\phi_0-\pi/2$. For $B<B^*$, we have $\delta=0$; while for
$B>B^*$, $\delta$ is a function of $z$ as shown in
Fig.~\ref{delta}.

\begin{figure}[ptb]
\begin{center}
\includegraphics[
width=6.in ]{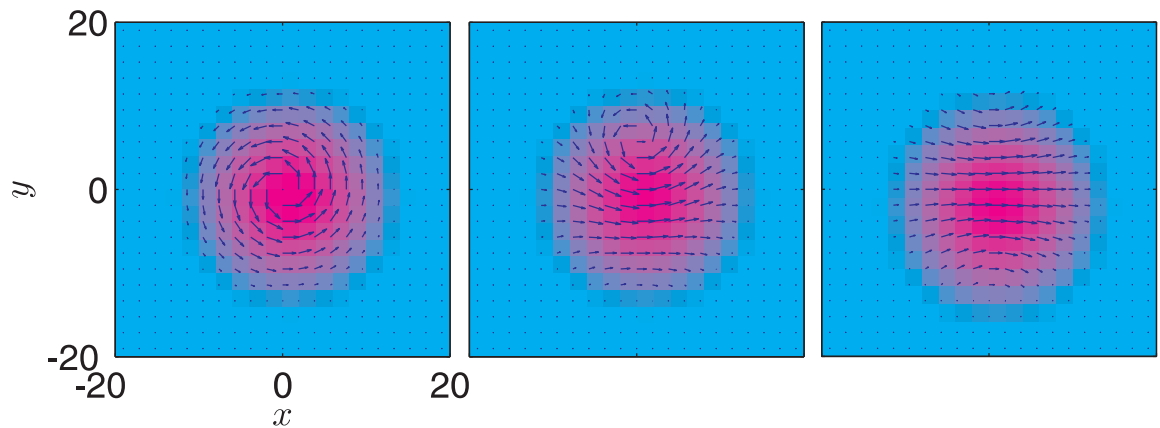}
\end{center}
\caption{Magnetic vortex under a transverse field for $N=1.5\times
10^6$, $q=0.5$, $\lambda=4$, and, from left to right, the
corresponding magnetic fields are $B_x=0$, $0.0143$, $0.0286$, and
$0.0429$ in units of mG. Shown are the spin structures in the
$z=0$ plane.} \label{mvorbx}
\end{figure}

{\em Transverse field} --- Now we consider the effect of a uniform
transverse field along the $x$-axis. As the field strength $B_x$
increases, similar to what happens in nanomagnets \cite{heng}, the
spin vortex starts to move away from the center and perpendicular
to the applied field (along the $y$-axis), and eventually moves
out of the cloud. This is illustrated in Fig.~\ref{mvorbx}.

\section{Conclusion}
To conclude, we have shown that dipolar interaction plays a
crucial role in spinor condensates. We have provided here a
detailed study of the dipolar spin-1 condensate both within and
beyond the single mode approximation. In the SMA regime, the
system is analogous to the single-domain state of a magnetic
material with the spins uniformly oriented. For sufficiently large
dipolar interaction strength, the SMA becomes invalid, and spin
textures develop as a result of the interplay between the exchange
and the dipolar interaction. We have found that a pancake-shaped
condensate is particularly sensitive to the dipolar interaction.
In such systems, the spin texture takes the form of a spin vortex
which is analogous to the magnetic vortex observed in thin
magnetic films.

We have emphasized the connection between dipolar spinor
condensates and ferromagnetic materials. We should also point out
that there exist important differences between the two: While the
latter represent a classically ordered system, the former are
intrinsically quantum mechanical, as Bose condensates are
macroscopic quantum objects described by quantum mechanical
wavefunctions. Indeed we have seen that the spin vortex of a
spinor condensate is intimately coupled with the topological
charge of the condensate wavefunction: the spin components possess
quantized vorticity or persistent currents. The fact that the spin
textures and the motion of the atoms are intimately connected is a
manifestation of an important property of the dipolar interaction,
that is, the dipolar interaction provides a {\em spin-orbit
coupling}. This can be more clearly seen if we notice that the
dipole-dipole interaction potential (\ref{ddp}) transforms as
spherical tensors of rank 2 in both coordinate and spin space, and
may be cast into the following form:
\[V_{\rm dd}({\bf r},{\bf r}') = -\sqrt{\frac{6}{5\pi}}
\frac{\mu_0 g_F^2 \mu_B^2}{|{\bf r}-{\bf r}'|^3}\,\sum_{m=-2}^2
\,Y^*_{2,m}({\bf e})\,\Xi_{2,m} \,,\] where $Y_{2,m}$ is a
spherical harmonic of rank 2 and $\Xi_{2,m}$ its counterpart in
spin space with components given by
\begin{eqnarray*}
\Xi_{2,0} &=& \sqrt{ \frac{1}{6}}\,({\bf F}_1 \cdot {\bf
F}_2-3F_{1z}F_{2z}) \,, \\
\Xi_{2,\pm 1} &=& \pm \frac{1}{2}\,(F_{1z} F_{2\pm} + F_{1\pm}
F_{2z})\,,\\
\Xi_{2,\pm 2} &=& -\frac{1}{2}\,F_{1\pm} F_{2\pm} \,.
\end{eqnarray*}
The dipolar interaction is invariant under simultaneous rotations
in coordinate and spin space and therefore conserves the total
angular momentum (spin $+$ orbital). However, it does not conserve
separately the spin and orbital angular momentum. The spin-orbit
coupling induced by dipolar interaction also creates the
opportunity for the transfer of angular momentum between the spin
and the orbital degrees of freedom under the constraint of total
angular momentum conservation. This could lead to the Einstein-de
Haas effect in the dynamical evolution of a spinor condensates
\cite{edh, edh1}. Spin-orbit coupling effect has also been
predicted in ferromagnetic nanostructures \cite{dot}.

Finally, we want to mention that the Berkeley group led by
Stamper-Kurn have developed a wonderful technique capable of
non-destrctive {\em in situ} imaging of local spin textures in
atomic condensates \cite{imag}. Using this technique, they have
recently found tentative evidence of dipolar effects in spinor
condensate of $^{87}$Rb \cite{dan}. To summarize, dipolar spinor
condensates represent a novel class of anisotropic superfluid as
well as a new kind of magnetic material, many of its rich
properties are just starting to be explored.

\acknowledgments We acknowledge the financial support from US
National Science Foundation under the Grant No. PHY-0603475 (HP),
and from the National Science Foundation of China under Grant No.
10674141 and the ``Bairen'' program of CAS (SY).

\end{document}